# A Hybrid Recommender System for Recommending Smartphones to Prospective Customers

PRATIK K. BISWAS*

Artificial Intelligence and Data, Global Network and Technology, Verizon Communications, New Jersey, USA.

SONGLIN LIU

Artificial Intelligence and Data, Global Network and Technology, Verizon Communications, New Jersey, USA.

Recommender Systems are a subclass of machine learning systems that employ sophisticated information filtering strategies to reduce the search time and suggest the most relevant items to any particular user. Hybrid recommender systems combine multiple recommendation strategies in different ways to benefit from their complementary advantages. Some hybrid recommender systems have combined collaborative filtering and content-based approaches to build systems that are more robust. In this paper, we propose a hybrid recommender system, which combines Alternating Least Squares (ALS) based collaborative filtering with deep learning to enhance recommendation performance as well as overcome the limitations associated with the collaborative filtering approach, especially concerning its cold start problem. In essence, we use the outputs from ALS (collaborative filtering) to influence the recommendations from a Deep Neural Network (DNN), which combines characteristic, contextual, structural and sequential information, in a big data processing framework. We have conducted several experiments in testing the efficacy of the proposed hybrid architecture in recommending smartphones to prospective customers and compared its performance with other open-source recommenders. The results have shown that the proposed system has outperformed several existing hybrid recommender systems.

## 1 INTRODUCTION

During recent years, recommender or recommendation systems have come to play an unavoidable and a critical role in our daily lives, when it comes to online journeys. From retail, e-commerce, online advertisement, music streaming to video on demand, the use of recommender systems has become widespread and pervasive. With the ever-increasing volume of online information, recommender systems provide an effective strategy to overcome such information overload. These systems are applicable in scenarios where many users interact with many items. They can make recommendations about products, information, or services for users. Companies like Youtube, Amazon, Ebay, Netflix, Linkedin, Pandora, etc., use recommender systems to boost their revenues by helping users discover and purchase new items, through delightful and convenient user experiences.

Recommender systems suggest relevant items to users. Typically, they search and filter through large volumes of dynamically generated information to provide users with personalized content and services. They have the ability to predict whether a particular user would prefer an item or not based on the user's profile and/or his/her prior interactions with different items. A variety of techniques has been proposed over the years for performing recommendations, including collaborative filtering, content-based filtering, demographic filtering, etc. Hybrid recommenders combine some of these methods to improve recommendation performance and overcome the bottlenecks of individual techniques. Hybridization over collaborative filtering based recommendation is often a good choice to overcome the cold start problem associated with it and get the best of the advantages of the contributing techniques.

Ever since its introduction, deep learning has generated considerable interest in many research fields such as computer vision and natural language processing, owing not only to its stellar performance but also to its capability of learning feature representations from scratch. Recently, it has also started to flourish in such areas as information retrieval and recommender systems.

Deep learning models can be powerful in combining collaborative-filtering and content-based information. Deep learning can make collaborative filtering approach viable even for products with minimal customer interaction data. In this paper, we have presented a novel architecture for a hybrid recommender system to address the gaps and improve the accuracy of a stand-alone collaborative filtering system through deep learning. The proposed solution leverages external embeddings from Word2Vec and Universal Sentence Encoder (USE) for textual descriptions as well as embedded user and item latent features from collaborative filtering in deep learning. Furthermore, it views the input information from multiple perspectives and combines characteristic, contextual, interactive, structural and sequential information through a deep neural network in a big data framework. It alleviates the cold start problem by integrating side information about users and items. Our objective is to improve the overall quality of smartphone recommendations.

The remainder of this paper is organized as follows. In Section 2, we discuss related work. In Section 3, we highlight our main contributions. In Section 4, we introduce the concepts and terminologies. In Section 5, we define the problem statement and chronicle data preparation using a big data framework. In Section 6, we describe the proposed architecture of the hybrid recommender system in detail. In Section 7, we analyze the capabilities of the proposed model. In Section 8, we present the experiment, results and comparisons with other relevant open-source recommender systems. We conclude the paper in Section 9 with a summary and directions for our future work.

## 2 RELATED WORK

Related research can be grouped into three broad categories: 1) conventional recommender systems, 2) hybrid recommender systems, 3) deep learning based recommender systems.

Recommender systems [1-3] are software tools that enable users find and select products (items) from a given assortment. Deshpande et al. [4] defined recommender system as a *personalized information filtering technology used to either predict whether a particular user would like a particular item (prediction problem) or to identify a set of N items that would be of interest to a certain user (top-N recommendation problem)*. The scope, characteristics and benefits of recommendation systems have been discussed at length in [5-6]. Several blogs [7-10] are now available on the internet that provide a good overview on recommender systems in general and their associated concepts. In industry, recommender systems have become critical tools to enhance user experience and promote sales/services for many e-commerce [11] based online websites. One of the most successful companies is *Amazon*, which has spent over 10 years in developing its recommender system [12]. Amazon added about another 20% sales from recommendations in 2002. Another famous example is *Netflix*, an online movie rental company, which established the *Netflix Prize competition* [13] to improve the accuracy of its movie recommender system (*Cinematch*) by 10%.

Recent work has demonstrated a wide variety of efficient approaches for building recommendation systems, utilizing either collaborative filtering, content-based filtering, context-based filtering, knowledge-based inferencing or hybrid filtering [14-30]. *Collaborative filtering* (CF) [14-17] is considered as the most basic, mature, popular, and the easiest of methods to find recommendations based on historical reviews of user and item interactions. Simon Funk [18] published a detailed implementation of a regularized matrix factorization (MF) based collaborative filtering approach, known as "Funk SGD" in his blog. Paterek [19] introduced a bunch of novel techniques, including MF with biases, applying kernel ridge regression on the residual of MF, linear model for each item, and asymmetric factor models for collaborative filtering. Zhou et al. [20] and Fredrickson [21] have described the Alternating Least Squares (ALS) algorithm to address the performance bottlenecks associated with the Gradient Descent (SGD) and Singular Value Decomposition (SVD) based approaches for MF. Collaborative recommender systems have been implemented in different application areas. GroupLens [22-23] is a news-based architecture employing collaborative methods in assisting users to locate articles from massive news database. *Content-based filtering* (CBF) [24-26] recommends items based on the similarity between the preferences of the current user (properties of items already investigated by the user) and item properties extracted from item descriptions. Incorporating contextual information in a recommender system [27-28, 55] helps to get a clearer picture of the situation of any individual, place or object, which is of relevance to the system for prediction. Knowledge-based recommendation [29-30] attempts to suggest objects based on inferences about a user's needs and preferences. *Hybrid recommender systems*, which combine two or more filtering techniques in different ways to increase the accuracy and performance of recommender systems, have been proposed in [30-35]. These techniques combine multiple filtering approaches in order to harness their strengths while leveling out their corresponding weaknesses. One of the first was Fab [36], which incorporated a combination of CF, to find users having similar website preferences, with CBF to find websites with similar content. The work of Burke in [37] was one of the first qualitative surveys on hybrid recommender systems, which can be classified, based on their operations, into weighted hybrid, mixed hybrid, switching hybrid, feature-combination hybrid, cascade hybrid, feature-augmented hybrid and meta-level hybrid.

Currently, deep learning enjoys a massive following. Recent advances in deep learning based recommender systems have gained significant attention by overcoming obstacles of conventional models and achieving high recommendation quality. Deep learning (DL) is able to selectively capture the non-linear and non-trivial user-item relationships, and enable the codification of more complex abstractions through data representations in the higher layers. Betru et al. [38] introduced three deep learning based recommendation models in [39, 40, 41], while Zhang et al. [42] compiled a more comprehensive survey of the different deep learning based recommender systems. Le [43] provided a bird-eye's view on ten different categories of deep recommendation systems based on the types of deep learning technologies employed. Wang et al. [44] reviewed a deep learning based collaborative auto encoder; Cheng et al. [45] proposed an App. recommender system for Google Play with a wide & deep neural model, while Wu et al. [46] presented a non-parametric recommendation model built on Recurrent Neural Networks (RNNs). Gomez et al. [47] presented the Netflix recommender system, which recommends over 80% of the movies watched on Netflix. Davidson et al. & Covington et al. [48-49] described deep neural network (DNN) based recommendation algorithms for video recommendation on YouTube. Okura et al. [50] discussed a RNN based news recommender system for Yahoo News. Ilin et al. [51] described a deep learning powered recommender system (IKI Service) for feeding personalized contents based on user interests.

DNNRec [52] and DeepHCF [53] are two deep learning based hybrid recommender systems. DNNRec leverages embeddings, combines side information with a very deep network. DeepHCF uses two sources of data, i.e., interaction matrix and item reviews, to train two deep models, i.e., a Multi-Layer Perceptron (MLP) and a Convolutional Neural Network (CNN) via joint training separately on the two different data sources. DeepFM [54] is another end-to-end model that seamlessly integrates a factorization machine (for modeling low-order interactions) with a MLP (for modeling high-order feature interactions). ConTagNet [55] is a context-aware tag recommendation system, which learns image features by CNNs and context representations by a two-layer, fully connected feedforward neural network. DeepCoNN [56] adopts two parallel CNNs to model user behaviors and item properties from review texts, utilizing a word embedding technique to map the review texts into a lower-dimensional semantic space, while preserving the word sequence. NetEase [57] is a session-based recommendation model for a real-world e-commerce website, which integrates the RNN with a feed-forward network to predict what the user will buy next based on the click history. Wei et al. [58] have proposed a hybrid recommendation model, combining collaborative filtering with deep learning, to address the cold start problem.

## 3 MAIN CONTRIBUTIONS

In this paper, we have proposed an innovative architecture for a hybrid recommender for recommending smartphones to prospective and existing customers. Our main contributions and advantages can be summarized as follows:

1. We have combined ALS and DNN through a flexible, open-ended architecture that can support both regression and classification.
2. We have integrated characteristic, contextual, interactive, structural & sequential information and trained and validated our model in a big data framework, including Spark Cluster and GPU based nodes.
3. We have used CNN, LSTM (Long Short Term Memory) and MLP in the same architecture to explore different perspectives from the same devices.
4. We have made extensive use of pre-trained embedding models.
5. We have demonstrated that the proposed recommender can overcome the shortcomings of the traditional recommenders and even outperform several of the open-source, state-of the-art hybrid ones in recommending smartphones.

Our contributions make our recommender distinct and different from other hybrid recommenders.

## 4 PRELIMINARIES – CONCEPTS AND TERMINOLOGIES

In this section, we clarify key concepts and terminologies and explain certain technologies, with their strengths and weaknesses, which provide the foundation for our work.

Recommender systems function mainly with two kinds of information, namely, *user-item interactions*, e.g., ratings, scores, likes, etc., and *characteristic* (*content-specific*) information, e.g., user and item features (attributes), contextual features, etc. Based on this, there are four types of algorithms used in recommender systems, namely, *collaborative filtering*, *content-based filtering,* *context-based filtering* and *hybrid filtering*. Accordingly, there are mainly four types of recommender systems.

### 4.1 Collaborative Filtering

*Collaborative filtering* based recommender systems [14-23] use past user-item interactions to recommend new ones. They look at these interactions to detect similarities between users/items based on their estimated proximities. A *user-item interaction matrix*, also known as *utility matrix*, records past user-item interactions and drives the algorithm behind these systems. Collaborative filtering (CF) based methods are further sub-divided into two categories, namely, *memory-based* approaches [4, 6-7, 9, 16], which are based on *nearest neighbor search*, and *model-based* approaches [17-23], which use *matrix factorization* to learn user and item representations, also known as *embeddings* or *latent factors*, from the utility matrix.

*Memory-based* collaborative filtering algorithms [4, 6-7, 9] are of two kinds, namely, *user-user* and *item-item*. *User-user* algorithms [6-7, 9] find closest users and suggest most popular items amongst these users. They are "user-centered" as they represent users based on their interactions with items and evaluate distances between users. Each user is represented by its vector of interactions with different items, i.e., by one row in the user-item interaction matrix. These systems have low bias and high variance. *Item-item* algorithms [7, 9, 16], on the other hand, find items similar to the items that the user has interacted with. They are "item-centered" as they represent items based on the interactions that the various users have had with them and evaluate distances between those items. Each item is represented by its vector of interactions with every user, i.e., by one column in the user-item interaction matrix. These systems have high bias and low variance.

Model-based collaborative filtering algorithms [7, 9, 17-23] assume an underlying latent or hidden model behind user-item interactions and try to discover it to make new recommendations. They use matrix factorization to discover latent lower dimensional space of features to represent users and items, i.e., they attempt to reduce dimensionality of the interaction matrix and approximate it by two or more smaller matrices with fewer latent (hidden) components. Matrix Factorization (MF) algorithms decompose the sparse user-item interaction matrix into two smaller and dense matrices: a user-factor matrix (containing users' representations) and an item-factor matrix (containing items' representations), such that a dot product of the user-factor matrix with the transpose of the item-factor matrix produces the interaction matrix. Training loss can be reduced by comparing non-empty ratings to predicted ratings and regularized by adding a penalty term while keeping values of latent vectors low.

*Stochastic Gradient Descent*, *Singular Value Decomposition* and *Alternating Least Squares* are three common techniques used for MF. User-factor and Item-factor matrices can be learnt through any of these three methods. The most popular training algorithm [18] is Funk's Stochastic Gradient Descent (SGD) that minimizes loss by gradient updates of both columns and rows of user and item matrices, without loading all of the data at the same time in computer's memory. Singular Value Decomposition (SVD) is another matrix decomposition technique [19] that decomposes the interaction matrix into three matrices: user, non-negative diagonal (having singular values) and item. Although popular, these techniques are not very scalable. Alternatively, one can use the much faster Alternating Least Squares (ALS) method [20-21] that iteratively optimizes user and item matrices by the general least squares step. ALS can run in parallel and there are several ways to distribute the computation of ALS depending on how we distribute the data. Once the matrix is factorized, a user vector can be multiplied by any item vector for estimating the corresponding interaction rating/score. The resultant user-item interaction matrix, i.e., the dot product of the eventual (learnt) user-factor and the transpose of the item-factor matrix, can also make both user-user and item-item recommendations.

The main advantage of collaborative filtering based systems is that they require no information about users or items and are applicable in many situations, where user/item characteristic information is unavailable. Moreover, for a fixed set of users and items, recommendation becomes more accurate as more users interact with items and bring more information to the user-item interaction matrix. However, the main disadvantage of such systems is that they suffer from the "*cold-start*" problem. They cannot recommend anything to new users or recommend a new item to any user without having any information about new users or items. They need enough information about user-item interactions to function accurately and properly. New users/items cannot be added to the matrix without the availability of associated interactions. Furthermore, these systems are not *scalable* due to their inability to scale to much larger data sets when more and more users and items are added into the database.

### 4.2 Content-based Filtering

*Content-filtering* based recommender systems [24-26] use additional information about users and/or items. They try to build models based on available user and/or item features to explain the observed user-item interactions. The inputs to the model are the features and the response is the interaction. The models are more constrained as user and item features are given at the onset and not subject to change. They have a high bias and low variance. Content-based filtering (CBF) casts the recommendation problem into either a *classification* (classifies if a user likes an item or not) or a *regression* (predicts the score given by the user to an item) problem. They hypothesize that if a user were interested in an item in the past, he/she would be interested in it in the future. They can be "user-centered", "item-centered" or both.

Content-based methods are computationally fast and interpretable. They do not suffer from the "cold-start" problem as much as the collaborative-filtering based methods. New users and/or items can be easily described by their characteristics (content). Therefore, the systems can be easily adapted to new items or new users. However, some of the problems associated with content-based filtering techniques are limited content analysis, overspecialization and sparsity of data. One common problem is that new users, unless explicitly asked for information, may lack a defined profile. Another problem could arise from new users or items with previously unseen features, but this could happen mostly at the onset. Besides, these systems may not capture user-item interactions very well as inputs to the model. They may not perform as well when recommendation becomes a multi-class classification problem with a very large number of classes or categories.

### 4.3 Context-based Filtering

*Context-filtering* based recommender systems [27-28] use the context or situation about the user, item or interaction to give recommendations to users. They extend the user/item convention by incorporating contextual information into the system. *Context* refers to the time, location, area and environment of the user that define a user's situational awareness. Context-based recommender systems may access the situational information directly using various techniques (such as GPS) without bothering the user. Contextual factors can be dynamic or static, depending on whether they change with time or not.

### 4.4 Hybrid Filtering

*Hybrid-filtering* based (Hybrid) recommender systems [30-35] combine multiple recommendation strategies in different ways to benefit from the advantages of each. However, in this paper, we will be confining ourselves only to hybrid recommender systems that combine collaborative filtering based algorithms with content-based modeling. These systems use both types of information, *user-item interactions* as well *as users' and items' characteristics/features/attributes (content)*, to avoid problems generated when working with just one kind. They depend on content-based recommendations when a user/item has no or little activity/interaction, but can fall back upon collaborative filtering when user-item interactions are widely available. As they collect more data, they become more accurate.

*Neural nets* can also be used to build hybrid recommender systems as they can combine collaborative approaches with content-based ones. Deep learning based methods can be very powerful in this regard [48-58]. The *Youtube* recommender system [48-49] built deep learning based hybrid models to recommend users' videos to watch, given their previous activities (search queries and videos watched) and static information (gender, location, etc.). A hybrid recommender integrating deep learning with collaborative filtering can address the cold start problem associated with collaborative filtering [58].

### 4.5 Deep Learning Based Recommender Systems

*Deep learning* based recommender systems [38-50] have become very popular in recent times. Deep Learning (DL) systems learn from both unstructured and unlabeled data and make intelligent decisions on their own using *artificial neural networks*. A neural network can also decompose the user-item interaction matrix. The *auto-encoder* acts similarly to matrix factorization. Neural networks can be trained to predict or classify interactions based on user and item features. *Deep Neural Networks* (DNNs) with *multiple hidden layers* and nonlinearities are more powerful than *shallow* networks, but harder to train if there is not enough data. In reality, millions of items and billions of examples are available to train the network to make recommendations. Deep learning has several advantages over traditional content or collaborative filtering based approaches to recommender systems [42-43]. *Multi-Layer Perceptions* (MLPs) facilitate efficient feature representation and provide non-linear transformation between inputs and output. As compared to matrix factorization, deep learning can be very flexible in including various factors into modeling and creating different *embeddings* [8, 49-50], for such large-scale scenarios. It can generalize previously unseen feature interactions through low dimensional embeddings. *Auto Encoder* (AE) is an unsupervised model which attempts to reconstruct its input data in the output layer. Deep auto-encoders [44] can be used in collaborative filtering to project interaction vectors to a latent space, similar to matrix factorization. In *Collaborative Deep Learning* [40-41], auto-encoders are trained on item features together with matrix factorization of interaction matrix to combine both collaborative and content-based approaches. *Restricted Boltzmann Machine* (RBM) is a two layer neural network consisting of a visible layer and a hidden layer with no intra-layer communication [39]. It can be easily stacked to a deep net. Textual descriptions of items, captured through the embedding layer [50-51] of a deep neural net, can be useful in recommending items to users. Item features can also include high quality visual information. In some recommendation scenarios (e.g. related to products in fashion), visual similarity can be more crucial than textual descriptions. Incorporating images into the input data, in these cases, can significantly improve recommendation performances. Image embeddings extracted by deep convolutional networks can contain both high level (product type) and low level (style of textures) similarity of items. *Convolutional Neural Network* (CNN) is a feed-forward neural network, with convolution layers and pooling operations, which can capture global and local features from unstructured multi-media data, thus significantly enhancing the model's efficiency and accuracy. CNNs can extract features from images, text, audio or video [53, 55-56]. Deep neural nets can also predict next action based on ordered historical actions and content. *Recurrent neural network* (RNN) is useful for modeling sequential data using loops and memories to remember former computations [46, 57]. Variants of RNNs include *Long Short Term Memory* (LSTM) and *Gated Recurrent Units* (GRU), which overcome the vanishing gradient problem using *gates*.

## 4.6 Embeddings

Embeddings [8, 50] map "entities" to a latent space with complex and meaningful dimensions. In matrix factorization based collaborative filtering, users and items are mapped to this space by how strongly associated they are with each of the dimensions. Words can also be mapped into a shared latent space such that the meaning of the word can be represented geometrically. *Embedding layers* [50-51, 55-56] in neural networks can be used to deal with the sparse matrix problem in recommender systems. The Embedding layer is defined as the first hidden layer of a network. Deep learning models often offer an *embedding layer* that can be used on textual data. It requires that the input data be integer encoded, so that each word is represented by a unique integer. The Embedding layer can be initialized with random weights or loaded with a pre-trained word embedding model in a type of *transfer learning*. It can be applied as part of a deep learning model where the embedding is either learnt along with the model parameters or remains unchanged during the learning process.

## 5 PROBLEM STATEMENT AND DATASET PREPARATION

Customers flock to on-line purchases due to the assortment of choices that are available on the retailers' websites. However, many ecommerce platforms fail to sell through a high percentage of their merchandise due to poor browsing experience of the users. Customers can spend hours scrolling through hundreds, sometimes thousands of items of merchandise without finding an item they may like. Shoppers need to be provided suggestions based on their preferences/needs for a better shopping experience that also boosts sales and revenues. Here, we have formulated this problem for smartphone selection through one of Verizon's use cases.

### 5.1 Problem and Scope

We can define the problem statement as the following challenge.

*Given a set of visitors to the Verizon website with their device selections for their first lines, can we recommend smartphones for their consecutive lines?*

Visitor profile information, the first line selection data and the browsing history of the visitor would be available for making subsequent recommendations. The recommender would be suggesting top five recommended devices based on the device that the prospective customer has already added into the 'shopping cart' for the first line. This would also enable quicker order completion. Based on these recommendations, Verizon Gridwall would display the top three devices as shown in Figure 1. It would primarily target new customers to Verizon, i.e., prospects, when they revisit for multi-line orders after adding the first device. However, it should also be applicable to new prospects (without selection) and existing customers for the recommendation of smartphones.

For building a recommender system for subsequent orders from new and existing customers, we have considered a set of smartphones and a set of visitors who have reacted to these devices in some way on the "My Verizon" website. The underlying dataset consists of many database tables that contain information about the smartphones, the visitors, including prospects (potential customers) as well as the customers (existing), and the combined browsing and purchase history of prospects and customers. Note, that henceforth, we have used the terms "visitors" to refer to both prospects and customers who visit the "My Verizon" websites and "devices" to refer to smartphones.

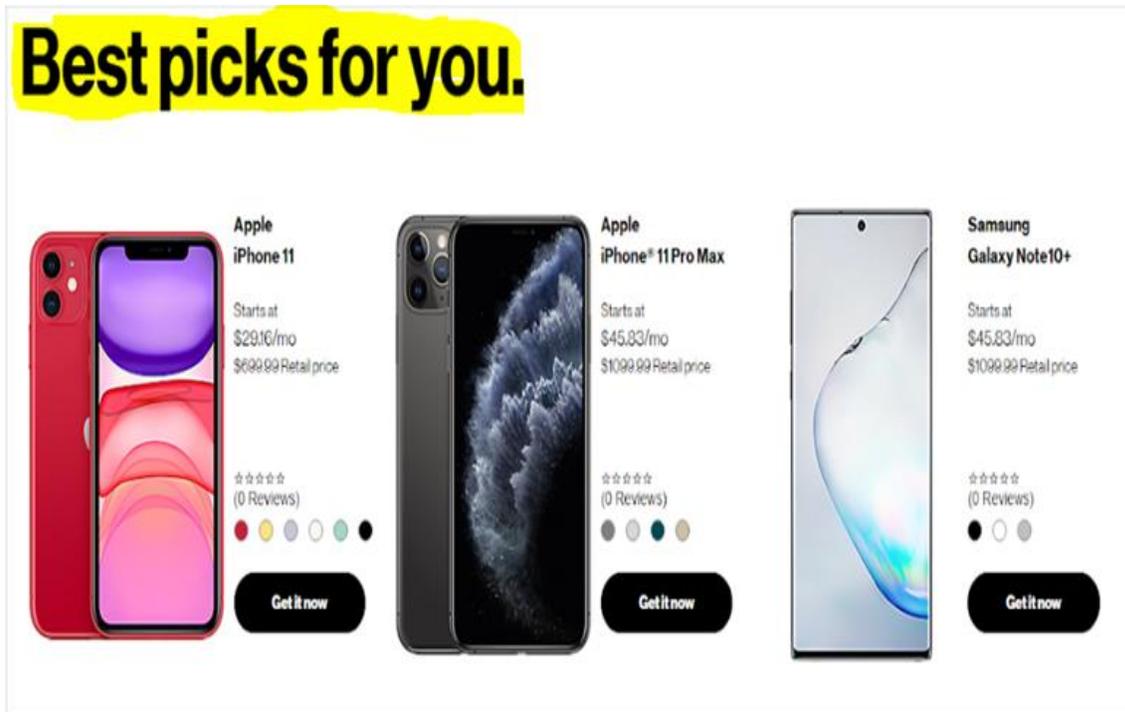

Figure 1: Top 3 Smartphone Recommendations to a Prospect

### 5.2 Data Collection

This is the first and most crucial step for building a recommender system. We have used *Spark cluster*, consisting of a *driver* and multiple *executors* (dynamically allocated), in the Verizon cloud, i.e., *Verizon Cluster Grid* (VCG), as the *big data framework* for building our recommender systems. The input data was extracted from 10 different Hive tables via *Spark SQL* in a *PySpark Jupyter notebook* based application. This data, pulled from multiple tables, was subsequently joined into a single *PySpark frame*.

Data about interactions can be collected in two ways: explicitly and implicitly. *Explicit* data is direct information about user's preferences provided intentionally, i.e. input from the users such as movie ratings, likes, etc. *Implicit* data contains indirect information based on un-intentional, sub-conscious user and item interactions.

### 5.3 Implicit Data

We have used *implicit* data in building our recommender systems. We have gathered this data from available data streams, capturing visitors' behavior through browsing history, search history, clicks and order history**.** We collected browsing and purchasing history for 90,615 visitors (prospects & customers) and 296 devices based on 955,193 interactions.

We have considered only three types of interaction events, namely, *view* (product list), *add to shopping cart* and *confirm order* or *purchase*. We have assigned a score to each interaction based on whether the smartphone was browsed (viewed), added to 'shopping cart' or ordered, using the *importance* of the interaction based on the ratio of distributions of the three interaction types. The *importance* of the interaction has been represented by a *confidence score* [17]. We have started out with missing values as negative preferences with low confidence scores and existing values as positive preferences with high confidence scores. We have assigned a score of *0.02* if a device is viewed, *0.04* if it is added to the shopping cart and *1.0* if it is ordered. Missing values have been assigned *0* by default. *More the interaction, more the confidence*. The *hits* (number of interactions) and the *average (Avg.) score* have further adjusted the confidence score.

### 5.4 Pre-processing and Feature Selection

Pre-processing of visitor and device data involved renaming & lowercasing of feature values (strings), transformation (casting) of feature types (e.g., integer to float), derivation of new features, checking for NULL categorical features, transfer of data from *PySpark frame* to *Pandas frame*, encoding of categorical features (e.g., label encoding), imputing missing values, scaling values for numerical features as well as aggregation of device interactions per visitor. Pre-processing has been followed by extensive data analysis and visualization (e.g., Figure 2 & Figure 3). The raw Hive tables contained 246 visitor and 188 device features. Feature selection reduced the feature set to 36 visitor and 18 device features. Out of the visitor features, 10 were contextual in nature tied to the visitor's browsing activity.

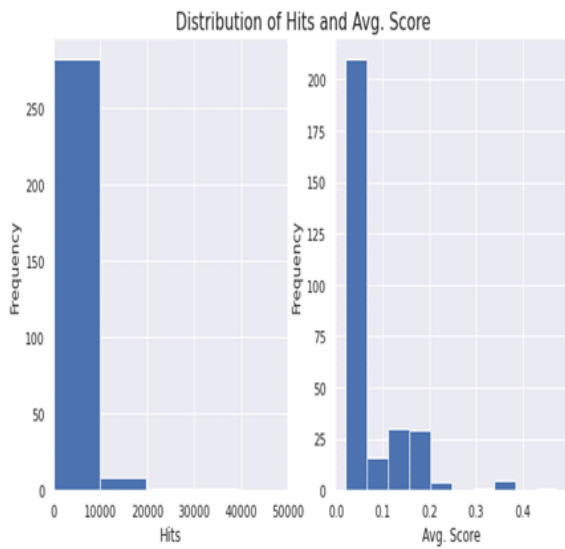
Figure 2: Histograms of Hits & Avg. Score

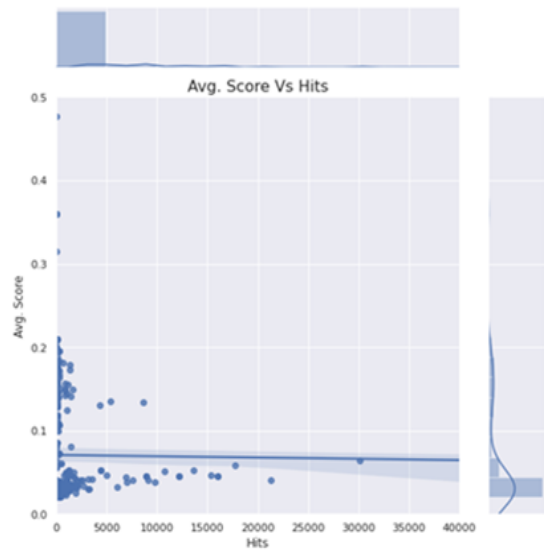
Figure 3: Dependence of Avg. Score on Hits

## 5.5 Data Reduction

The goal of a recommender system is to make recommendations that are relevant, up-to-date, popular, and useful. This stated objective can get adversely impacted if there are too many devices to choose from, many of which could be irrelevant. We have reduced the number of devices and visitors using *percentile* on hits, i.e., number of interactions between visitors and devices. We have used the 90th percentile value of 3225 on the number of interactions for a device as the cutoff threshold to filter out the devices (smartphones), which have a number of *hits* (interaction events) less than this threshold value. We ended up with 83,475 visitors, 30 devices and 823,257 interactions.

## 5.6 Interaction Score Weighting and Normalization

Using the raw confidence score as metric to make recommendations has its caveats. For one, it does not take the number of interactions, e.g., popularity into account. Secondly, this metric has the tendency to favor devices with a smaller number of interactions but with skewed and/or extremely high confidence scores. It is difficult to discern the quality of a smartphone with extremely few interactions. Considering these shortcomings, we have *weighted* the confidence scores, taking into account both the *average score* and the number of *hits*. The weighted scores have been further *normalized* for each visitor using the number of same and different devices the visitor has interacted with in the reduced input dataset. This has been done to keep the *cumulative* weighted score for the same device between 0 and 1 if the visitor has interacted multiple times with it. Figure 4 shows the *correlation* among the *hits*, *average score*, *weighted score and normalized score* in the reduced dataset.

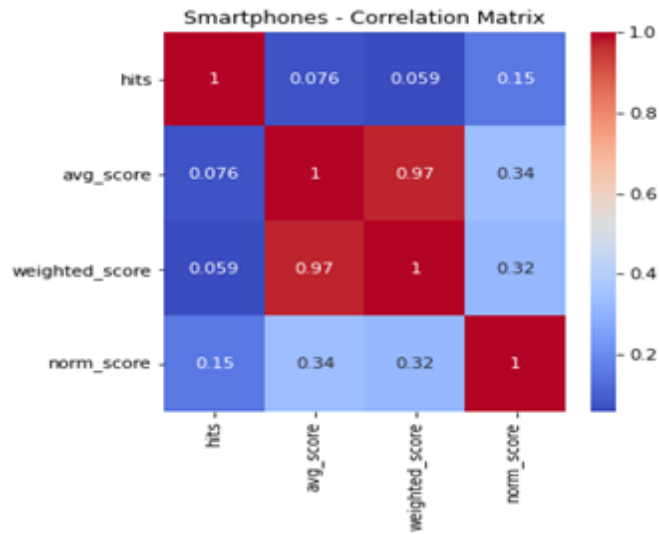

Figure 4: Correlations among Hits, Average Score, Weighted Score & Normalized Score

## 5.7 Input Formats

Data inputs to the recommender systems can be either *normalized* or *de-normalized*. Normalized input dataset has one row for *each* user-item interaction (e.g., Rating table). De-normalized input dataset has one row for each user with *all* his/her item interactions (e.g., User-item Interaction Matrix).

## 6 ARCHITECTURE FOR HYBRID RECOMMENDER SYSTEM

This section provides a description of a hybrid recommender that we are proposing for recommending smartphones. The recommender applies deep learning to overcome the shortcomings that collaborative filtering exhibits with sparse interaction data. The following are the main characteristics of the proposed model.

- It aggregates 4 self-contained neural networks (Figure 5)
- It integrates collaborative filtering with deep learning
  - It uses outputs from ALS as inputs to DNN model, implemented using Keras [60]
  - It can use either the visitor & device embeddings from ALS as inputs or the output interaction matrix from ALS as output (y) or both
- It models visitor behavior in combination with visitor and device (smartphone) characteristics
- It is context-aware
- It accounts for both structural and sequential nature of the input data
- It accepts de-normalized inputs
- It can support both regression and classification problems
- It executes on GPU based nodes

### 6.1 Architecture

The architecture of the proposed hybrid recommender is shown in Figure 5. The ensemble DNN can be used to address either a regression or a classification problem. The model consists of four parallel neural networks, i.e., N1, N2, N3, & N4, coupled deep down in a layer close to the output layer. With respect to Figure 5, this coupling is done at the "concatenate_2" layer which is the *shared layer*. It may be noted here that even though the outputs of N2 and N3 are shown to merge earlier, there is no layer at their merger and so they are meant to reach the *shared layer* separately. Now, three of these networks share the same input. N1 & N2 accept device identifiers, N3 accepts both device and visitor identifiers, while N4 accepts visitor, context and device features respectively as inputs to produce the corresponding confidence scores or the recommended classes as the model's output. Let $X = <X_1, X_2, X_3>$ be the input to the proposed model, where $X_1 = <x_{11}>$ represents the visitor identifier, $X_2 = <x_{21}, x_{22}, .., x_{2m}>$ represents an ordered sequence of $m$ device identifiers that $x_{11}$ has interacted with and $X_3 = <x_{31}, x_{32}, .., x_{3n}>$ represents $n$ associated visitor, context and device features (standardized and averaged).

Three separate pre-trained embedding models have been used in the proposed DNN architecture. At the *look-up (embedding)* layer, N1 & N2 use embedding matrices, populated with pre-trained Word2vec & Universal Sentence Encoder (USE) based embeddings, mapped from device names/descriptions (corresponding to device identifiers); while N3 uses device and visitor embeddings generated by the ALS.

N1 (Figure 5) is a self-contained *Convolutional Neural Network* (CNN), which consists of 3 blocks of *convolutional* layer, *batch normalization* layer and *max pooling layer*, followed by 1 *flatten* layer and 3 pairs of *dense & dropout* layers. It can be used to extract meaningful features from two-dimensional structures, created either by concatenating Word2Vec & USE generated embeddings from device descriptions or from digitized images of smartphones, if available (not shown in Figure 5). The output $y_{N1}$ from N1 is formulated by Eqn. (1).

$$y_{N1} = dropout(relu(dense(dropout(relu(dense(dropout(relu(dense(flatten$$
$$(maxpool2d(batchnormalization(relu(conv2d(maxpool2d(batchnormalization(relu(conv2d(maxpool2d$$
$$(batchnormalization(relu(conv2d(reshape(concatenate(emb\_3(X_2), emb\_4(X_2)))))))))))))))))))))))) \quad (1)$$

N2 (Figure 5) is a self-contained *Seq2Seq* network (S2S), which consists of 1 *lstm* layer, 2 *leakyReLU* layers, 1 *global max pooling* layer, 2 *dropout* layers and 1 *dense* layer. The Seq2Seq network is used with de-normalized inputs where each row in the input training set lists all the interactions of the corresponding visitor sequenced by time. It uses this input sequence to predict the next possible device for the visitor's interaction. The output $y_{N2}$ from N2 can be formulated by Eqn. (2) as below.

$$y_{N2} = dropout(leakyrelu(dense(globalmaxpool1d(dropout(leakyrelu(lstm(emb\_4(X_2)))))))) \quad (2)$$

N3 (Figure 5) is the self-contained *Neural Collaborative Filter* (NCF) that either takes a *dot* product or *concatenates* (not shown in Figure 5) of the ALS generated visitor and device embeddings and then passes the flattened result through 2 pairs of fully connected *dense* and *dropout* layers to uncover more nonlinear combinations from interactions. The output $y_{N3}$ from N3 can be formulated by Eqn. (3) as below.

$$y_{N3} = dropout(relu(dense(dropout(relu(dense(flatten(dot|concatenate(emb\_1(X_1), emb\_2(X_2)))))))))  \quad (3)$$

N4 (Figure 5) is a self-contained *Feature Processor* (FP), which passes a combination of visitor, context and device feature values to the shared layer through a pair of *dense* & *dropout* layers, after learning new features through nonlinear transformations. The output $y_{N4}$ from N4 can be formulated by Eqn. (4) as below.

$$y_{N4} = dropout(relu(dense(X_3))) \quad (4)$$

The *shared layer* concatenates results from all 4 networks before passing the resultant tensor to the *output layer* through a *dropout* and a *dense-dropout* pair. The output y from the model, for regression and classification, can be formulated by the following Eqns. (5) and (6).

Regression:

$$y = lambda(sigmoid(dense(dropout(relu(dense(dropout(concatenate(y_{N1}, y_{N2}, y_{N3}, y_{N4}))))))))  \quad (5)$$

Classification:

$$y = softmax(dense(dropout(relu(dense(dropout(concatenate(y_{N1}, y_{N2}, y_{N3}, y_{N4}))))))) \quad (6)$$

The lambda function is used for regression to guarantee that the predicted interaction score stays within the range defined by minimum and maximum values of interactions scores in the input data set.

### 6.2  Model Training and Validation

The model has been trained on de-normalized input dataset, where each row represented all interactions of each visitor, as its *input* and either the input or the output interaction matrix to/from ALS as its *output (target variable)*. For regression, the model takes each row of the target interaction matrix as it is, while for classification, it first converts each row of the interaction matrix into a one-hot vector (i.e., the maximum score is set to 1 and the rest to 0) before using it as the target variable. We have optimized the model using *Adam* optimizer over shuffled mini batches. We split both the input & output data 75% to 25% between training and validation sets. We prevented overfitting with dropout layers. For regression, we have used *MSE* as loss function and *RMSE* & *MAE* as the evaluation metrics, while for classification we have used *Categorical Crossentropy* as the loss function and *Precision*, *Recall, AUC, Accuracy* and *TopKCategoricalAccuracy* as the evaluation metrics.

Note that the number of blocks of *convolutional-batch normalization-max pooling layer*s in N1, the number of dense-dropout pairs in the various networks, the number of nodes in the dense layers, i.e., parameters to the model, and other hyper-parameter values have been optimized through train-test validation and cross-validation.

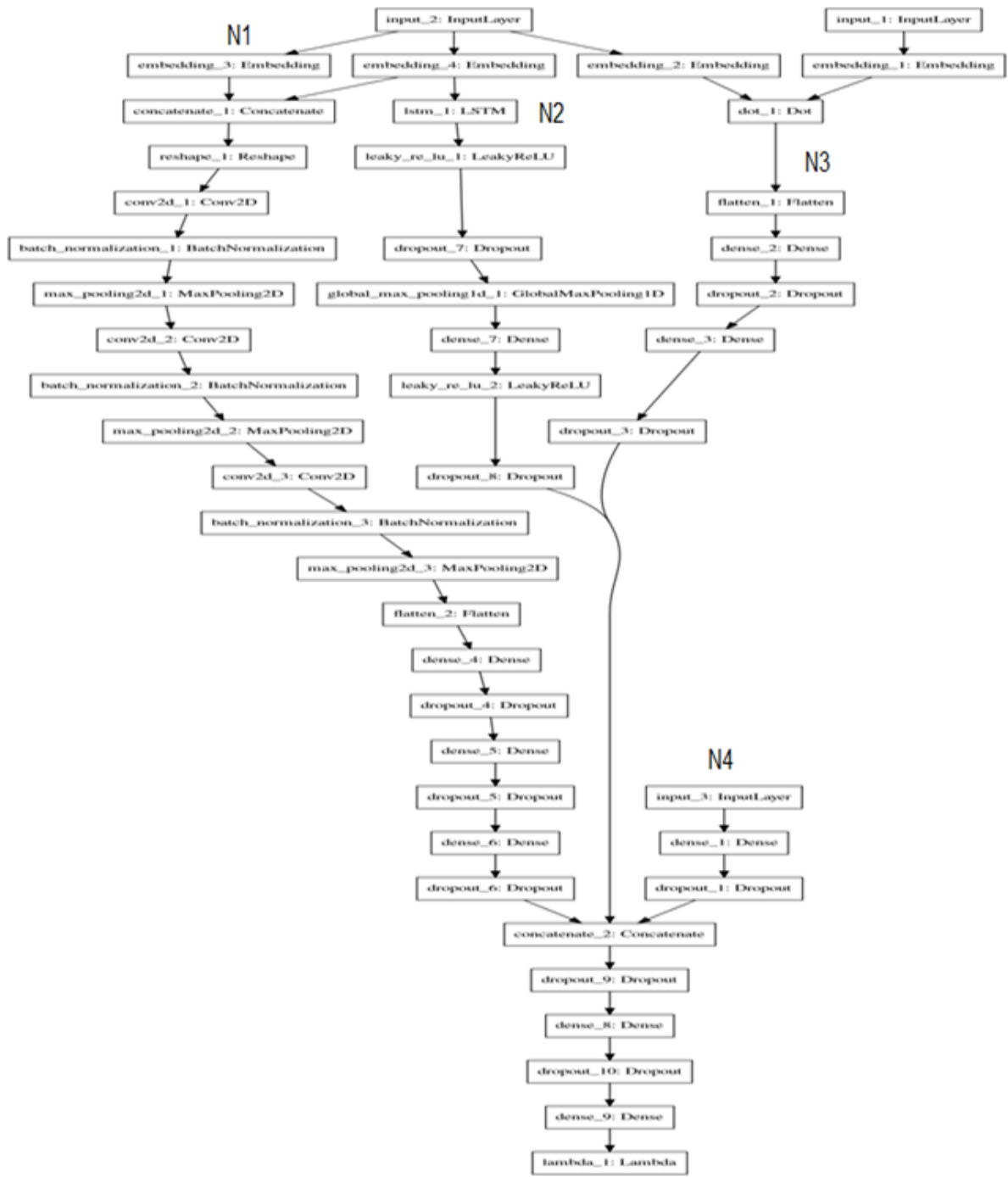

Figure 5: Architecture of the Proposed Hybrid Recommender

## 6.3 Cold Start Resolution

For prospects, the model uses available information from visitor profile and browsing context as inputs to N4. If the device information from the first line selection is also present, then it utilizes the corresponding embedding and the feature values. For unavailable visitor, context and device features, the model uses default values. The pre-trained embedding matrices in the lookup layers of N1 and N2 are provisioned with additional rows to account for unknown devices and visitors. For cold start situations, N3, *Neural Collaborative Filter* (NCF), doesn't have much role to play, but the ensemble of all 4 networks can still continue to make recommendations.

## 7 THEORY AND DISCUSSION

In this section, we present formal statements for our main claims with discussions on the capabilities of the proposed hybrid recommender.

***Definition 7.1:*** *Recommendation approach describes the way by which a broad class of solutions, namely, collaborative filtering, content filtering, hybrid filtering, etc., can be applied to the recommendation problem.*

***Definition 7.2:*** *Recommendation algorithm is the detailed strategy behind the recommendation approach.*

***Definition 7.3:*** *Recommender system is a process that uses a recommendation algorithm to predict user preferences.*

Some recommender systems can apply two or more recommendation approaches.

***Definition 7.4:*** *A hybrid recommender system (recommender) applies two or more recommendation approaches.*

***Definition 7.5:*** *Confidence score is the value that a recommender system assigns to a user's interaction with an item, representing the interaction's importance.*

For a visitor, *viewing* a smartphone has lesser importance than adding it to the *shopping cart*. If the visitor eventually confirms the *order*, i.e., purchases the smartphone, then the transaction event is complete, and the system can increase the confidence score to the maximum.

***Definition 7.6:*** *A weighted confidence score is the adjusted confidence value that takes into account the average score and the number of interactions for each item.*

***Definition 7.7:*** *A normalized weighted confidence score is the adjusted weighted confidence value for an item per user qualified by the number of times the user has interacted with the same and different items.*

This keeps the cumulative weighted score for the same device between 0 and 1 if the visitor has interacted multiple times with it.

***Definition 7.8:*** *The neighborhood ($N$) of a device is the set of all devices whose similarities to the given device are $\geq \lambda_N$, where $\lambda_N$ represents the similarity threshold value for neighborhood selection for that given device.*

Our recommender can generate an interaction matrix from a de-normalized input data set where there are as many rows as visitors and where each row includes all of the visitor's interactions with the various smartphones. This matrix can also yield a *smartphone-smartphone similarity* matrix, where any cell represents the similarity between any pair of smartphones. For each row or column in this matrix, the corresponding smartphone can then have a neighborhood ($N$) based on $\lambda_N$.

***Definition 7.9:*** *The recommendation coverage ($C$) for a user is the total number of items that are eligible for recommendation based on the chosen similarity threshold.*

Let the proposed recommender suggests $n$ smartphones to a visitor. Then the *recommendation coverage* for the visitor is the cardinality of the union of all the *neighborhoods* of the recommended smartphones, i.e., $C = |\bigcup_{i=1}^{n} N_i|$.

***Definition 7.10:*** *Representation Learning is a set of techniques for acquiring new knowledge from the raw data that makes it easier to extract useful information when building classifiers or other predictors.*

***Definition 7.11:*** *The flexibility of a recommender system refers to the openness of its architecture and its ability to change easily to suit different situations.*

***Definition 7.12:*** *The effectiveness of a recommender system refers to the degree to which it achieves its objective.*

In general, the objective of a recommender system is to provide relevant and useful recommendations that satisfy user needs. The effectiveness can be measured using different evaluation metrics.

***Claim 7.1:*** *The proposed hybrid recommender eliminates the bias that may be otherwise present in the positive browsing preferences with high confidence values.*

***Discussion:*** As discussed earlier, the assigned confidence scores do not take the popularity of the smartphones (Figure 2) into consideration. Recommendation becomes biased towards smartphones with fewer interactions but high confidence values. The system is predisposed to attach higher importance to a smartphone that has been purchased once, when compared to one that has been viewed thousands of times or added to the cart hundreds of times. As the number of interactions increases, the confidence score of a smartphone regularizes and approaches towards a value that is reflective of the device's quality and gives the visitor a much better idea of what to choose (Figure 3). The proposed recommender adjusts this bias in the confidence scores by weighting the confidence score of each interaction using the average score and the number of interactions (hits) for the corresponding smartphone (Figure 4).

***Claim 7.2:*** *The proposed hybrid recommender benefits from the use of word embeddings to exploit the semantic information from the smartphone descriptions.*

***Discussion:*** One of the drawbacks of the traditional recommenders is that they model users and items just based on the numeric values provided by users or user/item features and ignore the abundant information hidden in the textual descriptions. The proposed hybrid recommender maps product names/descriptions to word-embeddings using pre-trained deep models. In the look-up layer, it extracts the semantic information for each smartphone using a matrix of word/sentence embeddings, where the order of words is also preserved. Our results have shown that using word/sentence-embeddings for product descriptions can significantly improve the *effectiveness* of recommender systems.

***Claim 7.3:*** *The proposed hybrid recommender captures the complex non-linear visitor-device interaction patterns in the smartphone data.*

***Discussion:*** Conventional methods, such as matrix factorization and factorization machines use linear models to make recommendations. These models oversimplify the recommendation process and limit its expressiveness by assuming that the user-item interactions are linearly dependent on the user and item latent features. On the contrary, neural network based models can capture the non-linear interactions in the data with non-linear activations such as *ReLU*, Sigmoid, *Tanh*, etc. Neural networks can approximate any continuous function with arbitrary precision by varying the activation choices and combinations. This capability, as seen from Eqn.*(1)* - Eqn. *(6)*, has enabled our DNN based hybrid recommender to effectively capture the complex interaction patterns present in the input data, uncover non-linear relationships between features as well as from within interactions to precisely reflect the prospect's or customer's preferences.

*Claim 7.4: The proposed hybrid recommender improves representation learning for smartphone recommendations.*

*Discussion:* The proposed hybrid recommender attempts to provide better recommendations by learning the underlying explanatory factors and useful representations from the input data. It eliminates the need for handcrafted feature engineering by automatically learning visitor & device features and complex interactions. It processes heterogeneous content and context information such as text, images (if available), and numeric values for visitor and device attributes. It has shown potentials in learning representations from various sources.

*Claim 7.5: The proposed hybrid recommender mines the sequential structure in smartphone data.*

*Discussion:* Modeling sequential signals is an important topic for mining the temporal dynamics of visitor behavior and device evolution. LSTM achieves this with internal memory states and gates while CNN achieves this with filters sliding along with time. Both types of networks are part of the proposed architecture. The proposed recommender recommends smartphones to a visitor based on his/her browsing and/or purchase history, sequenced by hit time.

*Claim 7.6: The proposed hybrid recommender introduces high flexibility to smartphone recommendation.*

*Discussion:* The proposed hybrid architecture is modular in nature. It is easy to add new neural components to the existing architecture. It provides options to replace existing ones with others, e.g., *dot* with *concatenation,* one embedding matrix with another, etc. The number of *recurring blocks* that can be used in any of the component networks is also parameterized. It is also open-ended as the constituent DNN can be used for either regression or classification and can accept either the input interaction matrix to ALS or the output interaction matrix from ALS as its target variable to train and validate. Besides, the proposed recommender has been built using Keras [60], which is open-source and supports good modularization. Thus, we could easily build a hybrid and composite recommendation model to capture different characteristics and factors simultaneously.

*Claim 7.7: The proposed hybrid recommender overcomes the cold-start problem associated with collaborative filtering.*

*Discussion***:** Collaborative filtering based recommenders need enough information about user-item interactions to function accurately and properly. They cannot recommend anything to new users or recommend a new item to any user without having any information about new users or items in the user-item interaction matrix. As explained in Section 6.3, the proposed architecture of the hybrid recommender model easily overcomes this problem associated with collaborative filtering. In the cold-start scenario, only network N3, i.e., the *Neural Collaborative Filter* (NCF) doesn't contribute, but networks N1, N2 and N4 remain active, operate normally and can combine to recommend smartphones to prospective customers. If information on visitor, context and device features is completely unavailable, then the model uses default values. Furthermore, as stated earlier, the pre-trained embedding matrices in the lookup layers of N1 and N2 are equipped to handle unknown devices and visitors.

Next, we use **Definitions 7.8** & **7.9** to extend our *problem statement* (Section 5) and opine on its complexity, as in [59].

**Statement 7.1:** *Given $m$ smartphones with $m$ neighborhoods, a cold start prospect's device selection for the first line (if available), and a budget $k$; recommend $k$ smartphones for subsequent lines that maximize the recommendation coverage $C$.*

Let us name this problem as PDRC, i.e., Prospects Device Recommendation Coverage.

**Theorem 7.1**: *PDRC is NP-Hard.*

**Proof:** In order to prove that the search problem for PDRC is NP-Hard, we will perform a reduction from the known *Maximum Coverage Problem* (MCP), which is NP-Hard, to this problem. Given a family of sets $S = \{S_1, S_2, ..., S_m\}$ and a parameter $k$, MCP is to find $k$ sets such that the number of covered elements, i.e., $|\cup_{i=1}^{k} S_i|$, is maximized. Sets in $S$ may have some elements in common. Our reduction consists of demonstrating the following *two* steps in polynomial time.

Firstly, we will convert an instance of MCP to an instance of PDRC. An instance of MCP is a family of sets $S$ with $|S| = m$ and an integer $k$. We will consider $m$ smartphones and construct neighborhoods $N = \{N_1, N_2, ..., N_m\}$, by manipulating $\lambda_N$, such that $\cup_{i=1}^{m} S_i \leftrightarrow \cup_{i=1}^{m} N_i$ and $\forall i: 1 \leq i \leq m$, $|S_i| = |N_i|$. We will input these $N_i s$ to PDRC.

Secondly, we will show that a *yes* instance of MCP maps to a *yes* instance of PDRC and vice versa. Assume there exists a subset $S'$ of size $k$ in $S$, such that $|\cup_{i=1}^{k} S_i|$ is maximized. For every $S_j$ in $S'$, there is an equivalent $N_j$ in $N$. Thus, $S'$ yields $N'$ of size $k$, which then maximizes the *recommendation coverage* (C) for PDRC. Thus, $N'$ is a *yes* instance of PDRC and the proposed recommender can recommend the smartphones whose *neighborhoods* are included in $N'$. Now, assume that the proposed recommender has recommended $k$ smartphones and $N''$, with $|N''| = k$, is a subset of $N$ with $k$ neighborhoods that maximizes the *recommendation coverage* C. For every $N_j$ in $N''$, there is an equivalent $S_j$ in $S$. Thus, $N''$ results in $S''$. The neighborhoods in $N''$ form a subset $S''$ of $S$ of size $k$. Thus, $S''$ is a *yes* instance of MCP.

Therefore, MCP is reducible to PDRC. MCP is NP-Hard. Hence, PDRC is **NP-Hard.**

## 8 PERFORMANCE EVALUATION

We have considered various factors for the performance evaluation of our hybrid recommender. Online & offline methods, effectiveness (quality of recommendation), efficiency (training time), cold-start resolution, flexibility and performance comparison with other open-source, off-the-shelf packages are some of the considerations that helped us evaluate the performance of our hybrid model for recommending smartphones. We have looked at different possibilities of combining collaborative filtering with deep learning and have experimented with over 20 different architectures, ranging from simple to complex, using either only or a combination of Dense, CNN and LSTM layers, before deciding upon the final architecture presented in this paper.

### 8.1 Experimental Setup

We set up a Spark cluster, consisting a driver node and dynamically allocated, multiple executor nodes, on the GPU queue in the Verizon Cluster Grid (VCG) to conduct the experimentation with our proposed recommender system. The NVIDIA CUDA Deep Neural Network (cuDNN v7.6) accelerated our training process. We trained and tested all models on NVIDIA Tesla V100-SXM2-32GB GPU based nodes. The driver node used anywhere between 1 to 4 GPUs. We implemented our deep learning model in *Keras* [60], a well-known Python library for machine learning and deep learning. For ALS, we used the faster CPU based *Conjugate Gradient* method, and its *cython implementation* (CUDA version of ALS) from the *Implicit* [61] package as well as the implicit *ALS* module from *Spark MLlib* [62]. We trained and tested the proposed model on the smartphone dataset, described in Section 5. This dataset captured customer purchases between November 1$^{st}$ of 2019 and April 30$^{th}$ of 2020, together with their available browsing history going as far back as September of 2019. We evaluated the performance of the proposed model using both train-test validation as well as cross-validation. For train-test validation, we split the dataset 75:25. We optimized the hyper-parameters of the model using *Bayesopt* via cross-validation.

## 8.2 Evaluation Metrics

We evaluated our models for *efficiency* and *effectiveness*. The efficiency of hybrid recommenders is important to real world applications. We have compared the efficiency of different models by recording the training (as well as validation in some cases) time taken by each model to train on the same smartphone dataset. For ALS, we have also conducted the training across different platforms to measure the efficiencies of these algorithms vis-à-vis the platforms.

For *online* evaluation of the *effectiveness* of our model, we have compared top 5/10 recommendation of smartphones for some existing customers against their past conversion rates. For *off-line* methods, for measuring the *effectiveness* of our models and for comparing performances with other recommenders, we have primarily used the metrics, *RMSE* and *MAE* for regression and *Precision*, *Recall*, *AUC*, *Accuracy* & *TopKCategoricalAccuracy* for classification. For some off-the shelf recommenders, we have also looked at other metric like *MRR* (Mean Reciprocal Rank) if available.

## 8.3 Results and Model Comparisons

For collaborative filtering, we have benchmarked the *training time*, for different versions of ALS on the smartphone data, across multiple platforms. i.e., Spark cluster, CPU & GPUs in a big data framework. We have also compared the efficiency and effectiveness of ALS from *Implicit* [61] & *PySpark MLlib* [62] packages with various matrix factorization algorithms from *Surprise* [63] & *Turicreate* [64] packages. For overall performances of hybrid recommenders, we have compared the results from our deep learning model with those from other open-source packages like *Apple's Turicreate* [64], *Lyst's LightFM* [65] and *Marciej Kula's Spotlight* [66], using the same smartphone dataset.

Figure 6 and Figure 7 display benchmarks comparing the *training times* of various models of ALS, on CPU, GPU and Spark cluster, against the *number of factors* in the ALS model, using the smartphone data. Figure 6 shows that the *base* version on CPU (ALS-Base-CPU), the *conjugate gradient* version on CPU (ALS-CG-CPU) and the *conjugate gradient* versions on the GPU (ALS-GPU-1 & ALS-GPU-2) are 19-357x faster than the *cholesky* solver used in PySpark MLlib (ALS-Spark-MLlib). ALS-Base-CPU, ALS-CG-CPU, ALS-GPU-1 have also used 15 threads while ALS-GPU-2 used the native BLAS support, having disabled internal multi-threading. Figure 7 vividly shows that though there is little to choose among the non-spark versions of ALS, the ALS-CG-CPU and ALS-GPU-2 seem to perform a little better than the rest.

Table I compares the performances of the GPU (ALS-GPU-2) and *PySpark MLlib* versions of ALS with six other matrix factorization based recommenders from Scikit-learn's *Surprise* and Apple's *Turicreate* packages.

The results show that the GPU version of ALS (ALS-GPU-2) was the most *efficient*, while Turicreate's *Popularity* based recommender was the most *effective*.

Table II compares the performances of the proposed hybrid recommender against five other hybrid recommenders, from *Turicreate*, *LightFM* and *Spotlight* packages, on seven different evaluation metrics. The results establish that for regression (Table II – row 6) as well as classification (Table II – row 7), the proposed hybrid recommender *outperformed* all other recommenders in terms of *RMSE*, *Precision*, *Recall* and *AUC*, whenever results were comparable. Additionally, the proposed recommender classified each visitor's interactions for the given 30-class classification problem with 75% *Accuracy* (& 88% *TopKCategoricalAccuracy*). Moreover, Table I and Table II jointly prove that the proposed hybrid recommender, combining deep learning with collaborative filtering, significantly reduced the *RMSE* of ALS (ALS-GPU-2) from 0.19 to 0.0106. The "NAs" in Table I and Table II indicate that the corresponding metrics are either *not available* for the associated packages or *not applicable* to them (e.g., regression metrics are not applicable to classification models and vice versa).

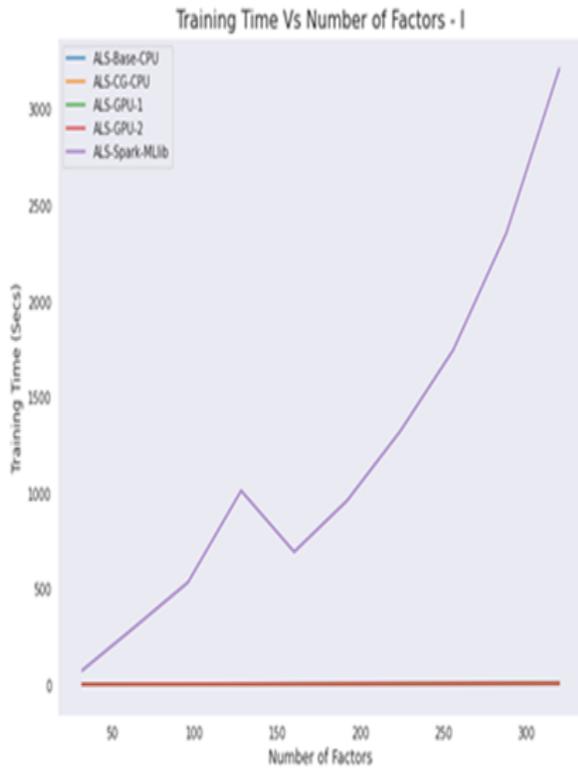 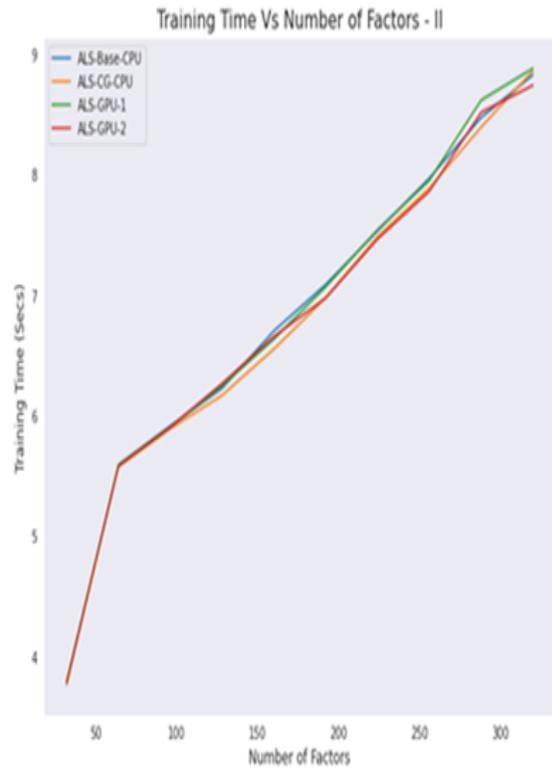

Figure 6: Efficiencies of ALS across Algorithms & Platforms    Figure 7: Efficiencies of non-Spark ALS Versions

COLLABORATIVE FILTERING ALGORITHM PERFORMANCE COMPARISONS

| No. | Package | Recommender | RMSE | MAE | Training & Validation Time (secs) |
|---|---|---|---|---|---|
| 1 | Implicit | ALS-GPU-2 | 0.1917 | 0.0436 | 3.97 |
| 2 | PySpark.ml | ALS | 0.0305 | 0.0199 | 20.85 |
| 3 | Surprise | SVD | 0.0221 | 0.0181 | 10.63 |
| 4 | | SVDpp | 0.0219 | 0.0178 | 55.70 |
| 5 | | NMF | 0.0236 | 0.0188 | 21.48 |
| 6 | Turicreate | Ranking Factorization (SGD) | 0.0277 | NA | 37.14 |
| 7 | | Popularity | 0.0185 | NA | 5.09 |
| 8 | | Item Similarity | 0.0319 | NA | 5.76 |

Table I: Evaluation Metric scores for Collaborative Filtering based Recommenders (NA stands for Not Available)

HYBRID RECOMMENDER PERFORMANCE COMPARISONS

| No. | Package | Recommender | RMSE | MRR | Precision | Recall | AUC | Accuracy (TopKCategoricalAccuracy) | Training & Validation Time (secs) |
|---|---|---|---|---|---|---|---|---|---|
| 1 | Turicreate | Item Content | 0.0279 | NA | 0.0552 | 0.2047 | NA | NA | 7.21 |
| 2 | | Factorization | 0.0151 | NA | 0.0599 | 0.2429 | NA | NA | 44.36 |
| 3 | | Ranking Factorization | 0.0188 | NA | 0.1128 | 0.4483 | NA | NA | 46.75 |
| 4 | LightFM | CF + Content | NA | 0.3115 | 0.1 | 0.38 | 0.6319 | NA | 359.69 |
| 5 | Spotlight | Explicit | 0.0894 | 0.1395 | 0.0525 | 0.3936 | NA | NA | 45.73 |
| 6 | Proposed | ALS + DL (DNN) | 0.0106 | NA | NA | NA | NA | NA | 1679.88 (20 epochs) |
| 7 | Proposed | ALS + DL (DNN) | NA | NA | 0.7952 | 0.4589 | 0.9307 | 0.7449 (0.8769) | 1618.17 (20 epochs) |

Table II: Evaluation Metric scores for Hybrid Recommenders (NA stands for Not Available)

## 9 CONCLUSION

In this paper, we have proposed a hybrid recommender system to overcome the shortcomings of stand-alone collaborative or content filtering based models and to achieve better performance in recommending smartphones to prospects and customers. The system combines the browsing history and purchase history of prospects and customers using Spark clusters on the Verizon cloud, and train an Alternating Least Squares (ALS) based collaborative filtering model & a deep learning based Deep Neural Network (DNN) model on GPUs, where the outputs from the ALS are fed sequentially to the DNN. The DNN is flexible in structure and consists of four components, namely, CNN, S2S, NCF and FP, which integrate at a shared common layer to model visitors and smartphones from a combination of structural, sequential, interactive, characteristic and contextual information and make recommendations. The proposed recommender leverages embeddings to exploit the semantic and visual information (if available) that exist in the associated textual descriptions and images of smartphones. It gains performance improvement from the following advantages: 1) combines structural, sequential, interactive, characteristic and contextual information in a big data framework; 2) uses pre-trained models for word/sentence embeddings; 3) learns both low- and high-order feature interactions; 4) introduces a sharing strategy of feature embedding to avoid large-scale feature engineering. We have conducted extensive experiments on the input data, using a big data framework, to compare the efficiency and effectiveness of the proposed hybrid recommender with several open-source, state-of-the-art models. Our experiment results show that the proposed model outperforms the open-source models in terms of the considered evaluation metrics.

We have built prototypes to demonstrate the efficacy of the proposed hybrid recommender to our customers. Currently efforts are underway to transition several of Verizon's existing content modeling and collaborative filtering based recommenders to the proposed framework.

**CONFLICT OF INTEREST STATEMENT**

The authors state that they are all employees of *Verizon* and this paper addresses work performed in course of authors' employment.